# Jason-RS, a Collaboration between Agents and an IoT Platform

Hantanirina Felixie RAFALIMANANA[1], Jean Luc RAZAFINDRAMINTSA[1], Sylvain CHERRIER[2], Thomas MAHATODY[1], Laurent GEORGE[2], Victor MANANTSOA[1]

[1] Laboratory for Mathematical and Computer Applied to the Development Systems, University of Fianarantsoa, Madagascar
[2] Université Paris-Est, Institut Gaspard Monge (LIGM),77454 Marne-la-Vallée Cedex 2
[1]`dhaliahfeli@gmail.com`
[1] `razafindramintsa.jeanluc@yahoo.fr`
[2]`sylvain.cherrier@u-pem.fr`
[1]`tsmahatody@gmail.com`
[2]`laurent.george@u-pem.fr`
[1]`vmanantsoa@moov.mg`

**Abstract.** In this article we start from the observation that REST services are the most used as tools of interoperability and orchestration in the Internet of Things (IoT). But REST does not make it possible to inject artificial intelligence into connected objects, ie it cannot allow autonomy and decision-making by the objects themselves. To define an intelligence to a connected object, one can use a Beleive Desire Intention agent (BDI an intelligent agent that adopts human behavior) such as Jason Agentspeak. But Jason AgentSpeak does not guarantee orchestration or choreography between connected objects. There are platforms for service orchestration and choreography in IoT, still the interconnection with artificial intelligence needs to be built. In this article, we propose a new approach called Jason-RS. It is a result of pairing Jason BDI agent with the web service technologies to exploit the agent capacity as a service, Jason-RS turn in Java SE and it does not need any middleware. The architecture that we propose allows to create the link between Artificial Intelligence and Services choreography to reduce human intervention in the service choreography. In order to validate the proposed approach, we have interconnected the Iot BeC[3] platform and the REST agent (Jason-RS). The decision-making faculty offered by Jason-RS is derived from the information sent by the objects according to the different methods of REST (GET, POST, PUT, and DELETE) that Jason-RS offers. As a result, the objects feed the inter-agent collaborations and decision-making inside the agent. Finally, we show that Jason-RS allows the Web of Objects to power complex systems such as an artificial intelligence responsible for processing data. This performance is promising.
**Keywords:** BDI Agent, BeC[3], Internet of Thing, Jason.

# 1   Introduction

The tendency towards the Web of Object (WoO) does not cease to increase from time to time. Given the number of objects to facilitate the daily life, the number of possible interactions and the applications that one can imagine become more powerful and rich. But what are the ways to make these application in a generic way? The object web orientation is considered as a great resource not only for managing objects [1] but also for questions of flexibility of exchanges. In the Web of Object (WoO) one is rather directed towards the use of Web Service [1].

We note that the most used Web Service type is REST[1] [2] [1] [3]. REST (Representational State Transfer) or RESTful is an architecture style for building applications (Web, Intranet, Web Service). It is a set of conventions, rules and good practices to respect and not a technology in its own right. Not only REST is stateless [4] (meaning that the service consumer does not have to store information about the way he is using the service); it is design to reduce the coupling between software pieces. In the Iot, REST may allow reconfiguration of the object, it reuse in a different application for a multiplatform application, REST service structure is easy to describe. REST allows machine to machine communication and is simplier to do than SOAP. The communication is done through the HTTP protocol. Nevertheless the possibility of using the resource via the web given by the REST, it still lacks intelligent data processing. Our contribution aim to add a complex data processing of the object web. We then opt for the reuse of our search fruit in [5] which we call Jason-RS or Jason-Rest for more precision.

Jason-RS gives the REST Web Service ability to the BDI agent Service [5]. Jason is a multi-agent platform based on the Agentspeak-L language [6] [7]. It is a platform for BDI agent, that is to say goal-oriented software agent, it is a smart agent. We have to look at the choice of the BDI agent because it is close to human behavior. [8] [9]. So the connexion of the latter with the REST Web Service allows him to publish his capacity as a Web Service. The processing of a complex action can be done inside Jason (prediction, composition of services, constraint logic programming) and could be published as a Web Service. Jason-RS can serve as an intelligent treatment center.

We focus in the collaboration between agent and Iot platform cause connected object had no intelligence, designed differently, so with this collaboration we can open in the intelligence of Iot with a sensor less, little human intervention. In our approach, we collect the information from the connected object and by Jason-RS we delegate the intelligent agent works. All change in the object behavior is sent to the agent BDI (as a perception) in order to provide a decision or to change the agent comportment. Through this perception that the agent can decide and delegate a task to agent or service choreography to object.

---

[1] https://www.supinfo.com/articles/single/5642-qu-est-ce-qu-une-api-rest-restful

The rest of this article is organized as follows: Section 2 will discuss the related work of our work. Section 3 describes in general the architecture of our contribution. In section 4 we will evaluate our work through an architecture. In section 5 we discuss a discussion and open challenge. We will conclude this article with a conclusion (section 6)

## 2      Related work

In this article we propose a new form of Web Service that the WoO can use. The use of REST in WoO is not an innovative work as researchers have already done research for the implementation of the latter to allow connected objects to trade [10] [11]. The Internet of Things is a system of physical objects that can be discovered, monitored, controlled or interact with electronic devices that communicate on various network interfaces and can potentially be connected to the wider Internet.

The webs of object is a concept say recent in computing that offers a future with which object of the daily life can integrate into the web [12]. The Web of Object is considered as a subset of the Internet of Things. WoO applies to standards and software infrastructures such as REST, HTTP and URIs to create applications and services that combine and interact with a variety of network devices. The key point of the WoO is that its implementation does not involve the reinvention of the means of communication given that they use the existing standards. It also gives an easy access to a wide population of programmer that used to handle such an architecture. It facilitates the integration of the IoT in the internet. It is also a solution that can open the connection of the Industry 4.0 [13].

View the exponential increase of the connected object, to reduce the cost, the required performance and to facilitate communication in the object internet, the W3C recently launched the working group on the Web of Objects. They aim to fight the fragmentation of IoT. They developed the initial standards for the Web of Objects [14]. To facilitate service choreography for Iot Cherrier describes in his thesis a new approach based on a platform to standardize the interoperability of heterogeneous objects [15].

Castellani and Al focused on the implementation of both CoAP and EXI technology for communication between objects by adopting the REST approach [16]. They implemented the CoAP protocol to enable communication between distributed connected objects. Then they also implemented the EXI library for compressing XML data into binary.

In the paper [2], they discuss how to create an event SOA. They found a new approach for coordinating IoT's services with the resource service. They named EDSOA the result of their research. It is an asynchronous SOA and responsive it can be considered as an event based service. We have inspired by this approach for the technical aspect of Jason-RS.

The Web of Object is used in almost any IoT field even in the field of health. On what [17] proposed a new instance of cloud server and set up data collection services on the heterogeneous devices they use. This aims to improve online health service by properly implementing the architectural web object principle for remote control.

Based on WoO, [4] describes the object-based web architecture that adopts the principle of Resource Oriented Architecture (ROA), hence the basis of RESTful. They implemented a smart gateway for smart meters. This allows the connected object management servers to keep track of the status and movements of the objects.[2] the HTTP protocol is used to facilitate access to the service. The answer emitted is in JSON format. The architecture proposed by [4] allows developers to use web programming languages such as (HTML, Python,...).

All of the research cited above are interesting from the point of view of exposure and communication capacity in the Internet of Things. Each with its strong point in communication and interconnection in the world of IoT. They all almost used the REST Web Service. But for us, a simple service is not enough not only to bathe the connected objects in the artificial intelligence but also to treat in intelligent way the data captured at the level of the objects. This treatment helps automatic decision-making thanks to the intelligent BDI agent that has the same reasoning mode as humans. We focus our research on the intelligent Web Service, that is, a Web Service associated with artificial intelligence in order to reduce human intervention to the orchestration of services in connected objects and decision-making. We have exploited here our previous research work, Jason-RS. An intelligent Web Service with BDI agents that perform the resolution and automation of certain tasks thanks to their ability to cooperate and exchange using the AgenSpeak (L) agent language. These agents may then exhibit functionality as a Web Rest Service. Jason-RS has the methods used in REST (GET, PUT, POST, and DELETE) to ensure the dialog with heterogeneous clients.

## 3     Proposed Architecture

### 3.1    General architecture

In this section we discuss the general architecture of our approach. As we are in the world of the IoT there is a mean of communication: the network. There are many types of networks today for the connectivity of objects such as wifi, Bluetooth for common devices, LoraWAN for IoT devices using long distance communication, and ZigBee, 6LowPan, EnOcean, Zwave for short range communication devices, etc. As connected objects do not have enough memory space and very limited energy to feed themselves permanently so it is important to outsource most of its function. Although the design of its objects is different (ie the objects are heterogeneous), a standardization platform is important to allow inter-object choreography. Taking into account all these parameters we have our control center server constituted by the Jason-RS server and the IoT server illustrated by Fig. 1. Generally the IoT server is a centralized platform of (Cloud, big data) whether it is an IoT server that administers the objects.

Jason plays a complex problem solver role as well as the complex system such as optimization ... The autonomous and goal-oriented capability through the BDI agent that it implements allows it to perform a task reasonably. This capability is now exploitable through a distributed component as a service [5] through Jason-RS. Then the objects can use it not only to send the sensor data but also to initiate a decision-making process at the agent level.

Communication between the IoT and the web is done through the HTTP protocol. With Jason-RS we can use all the existing methods of the REST and most importantly there is an interaction with the agent. As a result we can send data to the central control servers via the POST method. We can also retrieve the data or decision made by the GET method. JASON-RS recognizes directly the task that will send back to the agent.

At the agent level as soon as the service is called by the objects or other platform, the agent reacts and changes state in relation to the perception it receives from it environment.

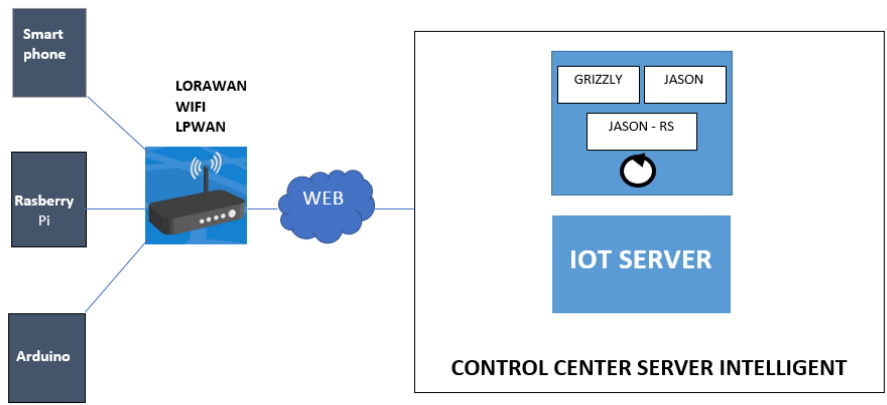

**Fig. 1.** General architecture of the JASON-RS implementation in the Web of Object

We have already studied in our article [5] the publication of Jason's BDI agent capability as a Web Service. We have proposed an approach allows deploying and running Web Services (REST and SOAP) pairs with a Jason BDI agent in java SE environment. We didn't use a modern Web-App server or application server, and we didn't developed a different middleware cause it requires a lot of time. Deploying an agent inside a server is a tedious task. Then in our strategy, we reused the existing Java frameworks called Non-Blocking Input Output. A small illustration for brightening the architecture is shown in Fig.2.

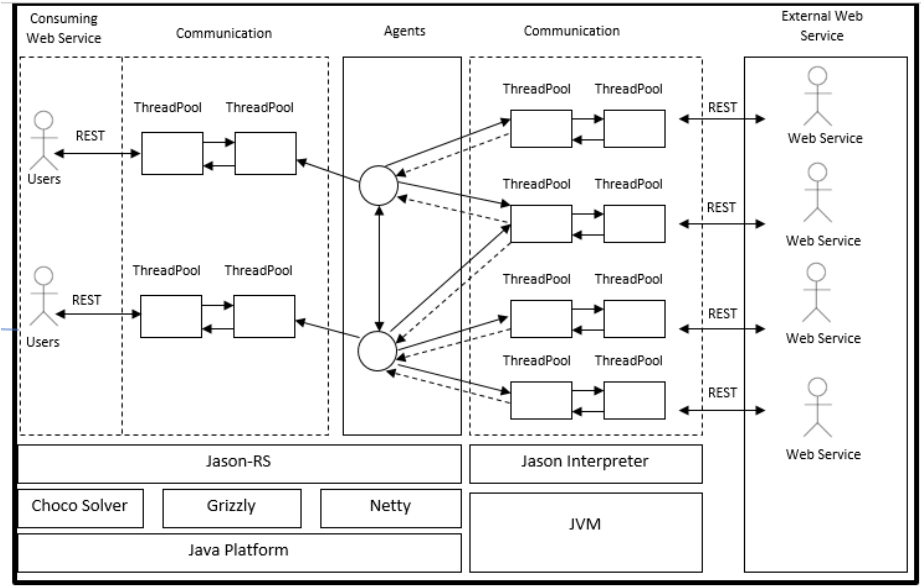

**Fig. 2.** General architecture of the JASON-RS micro-service [5]

Jason-RS offers an opportunity to exploit the world-class logic, which is embedded in Jason. It can associate them with other elements or heterogeneous application. When we can manipulate first-order logic, we can imagine a kind of possibility of complex problem solving as well as the common problem in artificial intelligence. What is interesting here is that the reasoning mode of Jason agent is near the human reasoning mode to achieve a goal. Jason-RS aims to outsource the capacity present in Jason as a service.

This initial Jason-RS that we have exploited with the collaboration with the Iot platform to intelligently process the data received from the connected object. By exploiting the horn clause and the first-order logic included in Jason we will use this ability to decide the action of objects. Then we take into account the behavior of the objects as perception for the agents.

To validate our approach we tested the interoperability of Jason-RS and heterogeneous objects choreographed with BeC$^3$.

### 3.2 Application of the architecture with a python object and BeC3

**Scenario.** To illustrate and motivate our approach, we introduce a concrete application using the following scenario. We take the example of the disposal of household waste in our paper [5]. Waste disposal agents have waste disposal costs that vary according to time constraints, people, distances traveled and evacuation loads. A decision-maker who gives these opinions to optimize the evacuation of waste thanks to the information that the evacuator agents gives him. Information of each of its agents could be changed thanks to the perception emitted by different connected objects. The Fig. 3 illustrates the application of this scenario. We used BeC$^3$ for the service Mashup. BeC$^3$ is a composer who allows a user to choreograph heterogeneous objects. In our experimentation we have created sensor objects choreographed in BeC$^3$ with other objects. This sensor sends the data captured via Jason-RS in POST for the agents to change the perception of the Jason agent (receiver agent). A decision-maker analyzes the present data and the cost of performing the tasks for each of his agents in order to decide which one is durable, and which one can perform a task at lower cost.

The sensor is controlled by the Python version of service compatible with BeC$^3$. When the service is launched, it can send a POST request to BeC$^3$ to identify itself and then another POST request to set up its feature handled by the device. Any mashup designed by a user can then be assured by a communication the service on each object. The decision made by the agent could be consumed via a GET request to the Jason-RS.

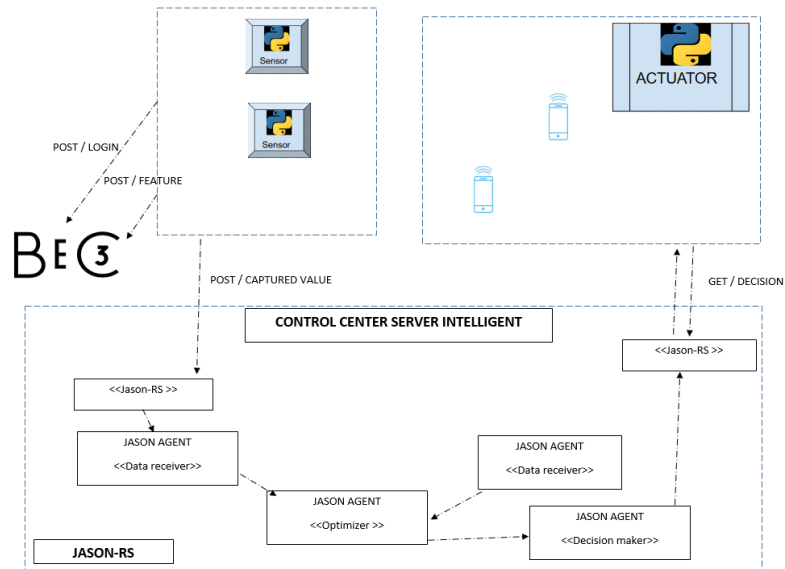

**Fig. 3.** Application interaction between object choreography and Jason-RS

## 4 Result and discussion

We tested this approach in two different objects: a python object, a mobile phone. We used Jason-RS to communicate the BeC$^3$ platform with choreographed objects and BDI agents who take the place of a decision maker instead of a human being (such as it is originally designed in BeC$^3$.

There is a demo script (in Python) APIDemo.py in Pylite project (resources dir). We have used xmpp[2] server for login, its specificity is the presence. So as we describe in the section 3 we must log in BeC$^3$ in order to create a virtual programmable object:

```
POST /login with content
{
  "username": {xmpp login},
  "password": {xmpp password},
  "service": {xmpp service (for example im.bec3.com)}
}
```

After login, create a BeC$^3$ object with the following JSON format:

```
POST /feature with content
{
    "name": "tot2",
    "path" : "truc/bidul21",
    "type": "switch",
    "details": "ooo",
    "widget": "none",
    "mqtt": false
}
```

Available types are accelerometer, button, buzzer, gauge, gps,
These object can be delete, update.

```
DELETE /feature/{id}
PUT /feature/{id}
{
  "data": {data (dependent of object type}
}
```

One object is associate with one BDI agent. To connect BeC$^3$ an URL is available from Jason-RS. To change the agent BDI perception a POST request is send by this url format:

```
POST jason_rs_based_url/object_agent/
{
  "data":value_from_python_object
}
```

To receive a decision from BDI agent, Jason-RS provide an URL that we can consume:
```
  GET jason_rs_based_url/agent_decider/decision
```

---

[2] https://xmpp.org/

The following sequence diagram presented in Fig.4 shows the exchange between these two platforms

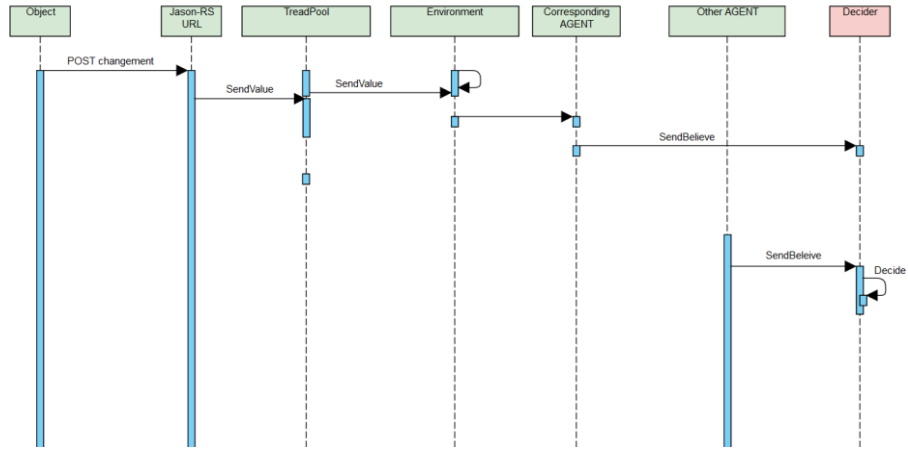

**Fig. 4.** Sequence diagram for communication between object a Jason agent by Jason-RS

We analyzed the performance of our approach via a browser.

**Table 1.** The calculation time of the exchanges between the objects and Jason-RS

| Device | Method REST | Task | Duration (millisecond) |
|---|---|---|---|
| Python Object | GET | Consumation WS | 200-450 |
| Smart Phone | POST, GET | Sending data | 428-1000 |

We tested our approach on both devices. During our experiment the return is almost <= 1000 ms. Our test area are done through the WIFI network so the discussion field is open for other types of networks available for the Internet of Things.

Agent-based internet of things is an interested field of research now [18], in the face of these works, we focus in the technical side of things, and we compare in table 2 what differentiates our approach compared to the state of the art.

**Table 2.** Comparison with a similar research domain work

| Work in field Jason agent and IoT | Agent flexibility in internet | Horn Logic | Taking in count Heterogeneity of IoT |
|---|---|---|---|
| [19] | ? | yes | ? |
| [20] | ? | ? | yes |
| [21] | ? | yes | ? |
| Jason-RS and IoT Platform | yes | yes | yes |

So in our approach we taking in count the heterogeneity of IoT by using a platform BeC$^3.$ To permit an automatic decision making we exploit the horn logic provided in Jason. This collaboration permit a flexibility in interacting between agent and heterogeneous connected object.

## 5   Conclusion

This article describes our contribution on service improvement in the world of Web of Object. We have integrated our previous Jason-RS research work into the sensor device connected to the web. Object of the IoT do not support the execution of artificial intelligence within themselve because there resources are very limited. Our approach is towards the decentralization of this AI treatment with Jason-RS. Not only is Jason-RS a Web Service but it has an artificial intelligence BDI engine running. So the complex data processing in the connected web object is provided by the Jason. The communication is done through HTTP protocol as it is more flexible and commonly adopted in the world of web. We tested our approach with a Mashup service platform for heterogeneous connected objects called BeC$^3$.Our approach therefore allows the object connected to the web to open up to data processing with artificial intelligence as well as complex systems itself.  As perspective we are considering the automatic injection of service inside the connected object to reduce the human task of defining the service choreography.